\documentclass[]{interact}

\usepackage{epstopdf}
\usepackage[caption=false]{subfig}
\usepackage{caption}
\usepackage[longnamesfirst,sort]{natbib}
\bibpunct[, ]{(}{)}{;}{a}{,}{,}
\renewcommand\bibfont{\fontsize{10}{12}\selectfont}

\theoremstyle{plain}

\theoremstyle{definition}

\theoremstyle{remark}

\begin{document}

\articletype{RESEARCH ARTICLE}

\title{A Novel Low Complexity High Resolution Spectrum Hole Detection Technique for Cognitive Radio}

\author{
\name{Sushmitha Sajeevu\textsuperscript{a}\thanks{CONTACT Sushmitha Sajeeevu Email: sushmitha\textunderscore p190077ec@nitc.ac.in} and Sakthivel Vellaisamy\textsuperscript{b}}
\affil{\textsuperscript{a}National Institute of Technology, Calicut, Kerala, India;  \textsuperscript{b}
National Institute of Technology, Calicut, Kerala, India;}}

\maketitle

\begin{abstract}
Cognitive radio is a potential solution to meet the upcoming spectrum crunch issue. In a cognitive radio, spectrum holes can be identified using spectrum sensing techniques. A high resolution spectrum hole detection can ensure even the smallest inactive portion in the spectrum is efficiently utilized. In this paper, a spectrum hole detection technique is proposed in which coarse sensing is done initially so as to detect occupied channels simultaneously. Spectrum holes in the occupied band can be efficiently detected using a fine sensing method. A two stage Frequency Response Masking (FRM) filter sandwiched between two Pascal structure based sampling rate converters results in arbitrary variation of bandwidth. This arbitrary variation of bandwidth can be utilized for fine sensing the spectrum such that the spectrum holes can be detected with high resolution. In the proposed method, high resolution in spectrum hole detection can be achieved without increasing the hardware complexity of the design. The hardware complexity of the proposed method is compared with the state of the art and is found to be significantly less.
\end{abstract}


\section{Introduction}
With the advances in wireless technology, use of wireless devices has elevated significantly.  A remarkable growth of wireless connected devices is expected in the near future, with the extensive adoption of Internet of things (IOT). Enormous amount of radio spectrum is required to support these wireless devices. But the available spectrum is scarce. Cognitive radio is a concept introduced to resolve the upcoming spectrum scarcity issue. With Cognitive radio, an unlicensed user, also known as secondary user (SU) is able to opportunistically or concurrently access spectrum bands owned by the licensed users also known as primary users (PU).\\
In a country, spectrum is allocated by Government agencies for defense, satellite operators, aviation, railways, service providers etc. Federal Communications Commission (FCC) in United States,  Telecom Regularity Authority in India, the Infocomm Development Authority (IDA) in Singapore, the Office of Communications (Ofcom) in the United Kingdom (UK) etc are the authorities allocating spectrum in respective countries [1]. Conventionally, different parts of the radio spectrum with certain bandwidth is allocated to different services. With such an allocation only the licensed user can utilize the spectrum assigned to that particular service irrespective of whether the licensed user is active or not.  Even though this particular allocation reduces the interference among different services, efficient utilization of spectrum cannot be achieved through this conventional allocation. Studies conducted worldwide such as Singapore, US, New Zealand, Germany and China have revealed that a huge amount of the allocated spectrum is underutilized [2]. It can be inferred that inefficient utilization of the spectrum contributes to the spectrum scarcity issue even more than the physical shortage of the spectrum.\\ Cognitive radio can offer an efficient communication with proper integration of the radio spectrum. In a cognitive radio, spectrum sharing schemes can be classified into different categories. Spectral sharing in a cognitive radio can be classified as centralized spectrum sharing and distributed spectrum sharing [3]. In centralized spectrum sharing, there will be a centralized entity which controls the spectrum allocation procedure. In distributed spectrum sharing, there will be no centralized operation, each and every node is responsible for spectral allocation. Based on the access behaviour, spectrum sharing can be classified into cooperative spectrum sharing and non-cooperative spectrum sharing. In cooperative spectrum sharing, interference information is shared between nodes and the spectrum is accessed in a cooperative fashion. Unlike the cooperative spectrum sharing, in non-cooperative spectrum sharing, nodes will not share information with other nodes and hence non-cooperative spectrum sharing will not result in efficient spectrum utilization. Based on the spectrum access technique, spectrum sharing can be classified into mainly three types; underlay, overlay and interweave.  \\
In underlay spectrum sharing technique [4, 5], secondary users access the channel concurrently with the primary user as long as the interference is below an acceptable level. The transmit power of the secondary user is limited by the interference constraint. In this technique, the secondary user need not wait for the transmission until the primary user is inactive and channel is idle. However this paradigm has increased complexity which is the main disadvantage of this spectrum access technique.  In overlay spectrum sharing technique [6], secondary user can concurrently transmit with the primary user however the secondary user must have complete knowledge of the primary user. In interweave spectrum sharing technique, transmission will be interference free. The basic principle of interweave spectrum sharing technique is listen and talk. Here the secondary users utilizes spectrum holes and  end their transmissions when the sensing algorithms indicate that primary users are resuming.  The importance of spectrum sensing comes in this scenario.\\
Spectrum sensing is the process through which the status of the spectrum is checked. If only one frequency band is taken into consideration it is called as narrowband spectrum sensing [7] whereas if multiple frequency bands are taken into consideration, it is called as wideband spectrum sensing. Matched filter detection [8], cyclo-stationary feature detection [9], covariance based detection [10], machine learning based spectrum sensing [11], energy detection [12] etc. are some of the narrowband spectrum sensing techniques. Optimal sensing can be achieved through matched filter detection even at low SNR region. Prior knowledge of the user signal is required for performing this detection but it is not always available. Hence matched filter detection is not always a practical solution. Cyclo-stationary feature detection can distinguish between signal and noise and it has got decreased probability of false alarm at low SNR environment. Meanwhile large sensing time is required to achieve good performance in this method. On the other hand, prior knowledge of the primary user signal and noise is not required in covariance based detection. But the computational complexity of this method is very high. If trained correctly, machine learning based spectrum sensing is a good approach. But this method requires complex techniques and large data set. Energy detection based spectrum sensing is the simplest spectrum sensing scheme in the literature. It has low complexity when compared to other spectrum sensing schemes.  Prior knowledge of the licensed user characteristics is not required in this method which resulted in the wide acceptance of this method. The energy consumption of a cognitive radio depends on the complexity of the spectrum sensing scheme adopted. In battery powered cognitive radios, low energy consumption spectrum sensing schemes are preferred. Hence energy detection based cognitive radio is an ideal choice in power efficient cognitive radio systems [13]. Wideband spectrum sensing (WSS) can be achieved by splitting the entire bandwidth into different frequency bands and sensing each frequency band using narrowband spectrum sensing techniques.\\
Using a spectrum sensing scheme, we can sense a band is occupied or not. Numerous filter bank based techniques [14],[15] were introduced to efficiently sense different bands present in a bandwidth. But the problem with filter bank based methods is that complexity of the design increases considerably when the number of bands increases. Hence proper sensing of narrow unoccupied bands fails using filter bank based methods. Later in [16,17], introduced a design in which the unoccupied portions of an occupied band can be sensed. Here in this design,  without increasing the number of bands fine sensing can be achieved. A variable bandwidth filter was used for the sake of fine sensing in this method. The resolution to which fine sensing is done is directly proportional to the complexity of the design in [16,17].\\
This paper proposes a novel spectrum sensing technique using Pascal structure based continuous variable bandwidth filter and energy detection technique. In this design, high resolution in spectrum sensing can be achieved without increasing the complexity of the design.\\
The paper is organized as given below. Section 2 briefly explains the Pascal structure based continuously variable bandwidth filter. Section 3 explains the proposed method for spectrum hole detection. Section 4 includes the design examples and result analysis. Section 5 and 6 describes hardware complexity calculation and evaluation respectively of the proposed scheme. Section 7 concludes the paper.
\section{Pascal Structure based continuously variable bandwidth filter}
Arbitrary variation of bandwidth can be achieved using Pascal structure based continuously variable bandwidth filter [18]. This can be used to get tunable frequency responses with any arbitrary bandwidth.
\subsection{Basic Pascal structure}
\begin{figure}[h]
\centerline{\includegraphics[width =10cm]{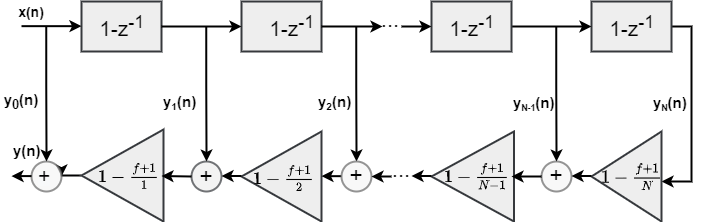}}
 \caption{Pascal Structure [19]}
 \label{Fig. 1}
 \end{figure}
Pascal structure is a variable fractional delay filter introduced in [19].
 The transfer function of the  fractional delay filter can be represented by the equation
\begin{equation}
H(z,f)= \sum_{k=0}^{N}P(f,k)(1-z^{-1})^{k} 
\end{equation}
where, \(H(z,f)\) is the transfer function of Pascal structure, \(f\) is the fractional delay parameter,  \(P(f,k)\) is the Pascal polynomial of k-th degree.
\begin{flushleft}
 \(P(f,k)\) can be written as   
\end{flushleft}
\begin{equation}
   P(f,k)=(1-\frac{f+1}{1})(1-\frac{f+1}{2})(1-\frac{f+1}{3})...(1-\frac{f+1}{k})
  \end{equation}
 The implementation of equation 1 is shown in Figure 1. Low hardware complexity when compared with the state of the art  [20],[21] and high modularity are the special features of the Pascal structure.

 \subsection{Continuously Variable Bandwidth digital filter}
 \begin{figure}[h]
\centerline{\includegraphics[width=9 cm]{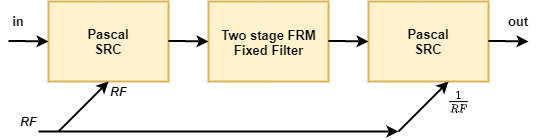}}
\caption{Block diagram of the Pascal structure based continuously variable bandwidth FIR filter}
 \label{Fig. 9}
  \end{figure}
  Figure 2 shows the basic block diagram of the Pascal structure based continuously variable bandwidth FIR filter. It is through the Pascal structure based sampling rate converter, the continuous variation of bandwidth is achieved. There is a fixed filter which is designed using two stage FRM fixed filter in which Parks McClellan (PM) algorithm is employed. The design of Pascal structure based sampling rate converter is explained in section 2.3. 
  \begin{equation}
   RF= \frac{BW_{Orig}}{BW_{Des}}
  \end{equation}
  An arbitrary bandwidth (\(BW_{Des}\)) can be obtained from the original fixed filter bandwidth (\(BW_{Orig}\)) using the above equation. The RF (Reduction factor) value has to be adjusted so as to get desirable bandwidth from original bandwidth. In Figure 2, the input is an impulse function and the output is the filter with bandwidth (\(BW_{Des}\)).
  \subsection{Design of Sampling rate converter using Pascal structure} \label{Section 4}

 \begin{figure}[h]
 \centerline{\includegraphics[width=8.5 cm]{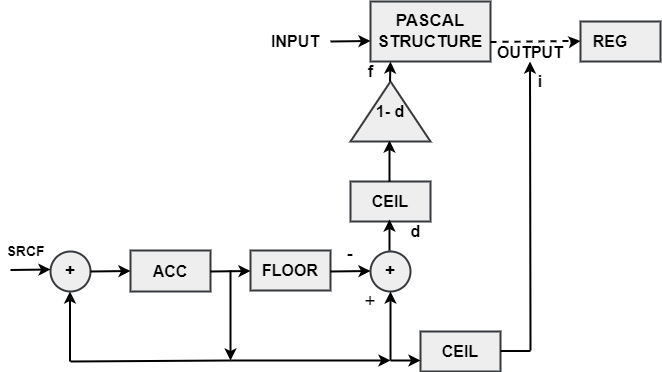}}
 \caption{Proposed Sampling rate converter (SRC)}
 \label{Fig. 5}
 \end{figure}
 Sampling rate conversion is a technique for converting the initial discrete time samples for the same continuous time signal to a different set of time samples.In order to delay the input signal by any desired amount of time, fractional delay structures are used. As a result, it is possible to determine the samples at non-integer multiples of the sampling period.  Pascal structure is a fractional delay structure that has high modularity and low complexity. Figure 3 shows the proposed design of a sampling rate converter using Pascal structure. The Sampling Rate Converter's operation is described in the following steps.
\begin{enumerate}
    \item The index values for each sampling rate conversion are the values in the accumulator. The accumulator (ACC) is first loaded with zero.
    \item The output pointer (dotted lines) points to the first location (n=0) of the output register (REG).
    \item The fractional part (d) of the accumulator is separated from accumulator value.
    \begin{equation}
    d= ACC-\lfloor{ACC}\rfloor
    \end{equation}
    \item The fractional delay (f) of the Pascal structure is calculated by
    
\begin{equation}
f=1-d  \iff d\neq0
\end{equation}
\begin{equation}
f=d    \iff d=0
\end{equation} 
 The ceil and the multiplier block having (1-d) is meant to perform the equations 5 and 6.
\item The OUTPUT of the Pascal structure corresponding to the fractional delay (f) is calculated.
   
    \item The ceil of the value in the accumulator,  gives an index(i) value.
    \begin{equation}
    i=\lceil{ACC}\rceil
    \end{equation}
    \item The value at the \(i^{th}\) index of the Pascal structure OUTPUT is saved to the pointed location of the register (REG).
    \item The output pointer (dotted lines) points to the next location of the  register (REG).
    \item Accumulator is accumulated with sampling rate conversion factor( SRCF) value.
    \begin{equation}
    ACC = ACC + SRCF
    \end{equation}
    \item Repeat steps 3-9 until the value in the accumulator is not greater than \((L_{I}-1)\), where \(L_{I}\) is the length of the input signal (The signal to be sampling rate converted).
\end{enumerate}
 
 The Pascal structure based sampling rate converter can be used for continuously varying bandwidth digital filters. This continuously varying bandwidth digital filter can be efficiently used in the context of spectrum hole detection.

\section{Proposed Method for spectrum hole detection}
 \begin{figure}[h!]
 \centerline{\includegraphics[width=14 cm]{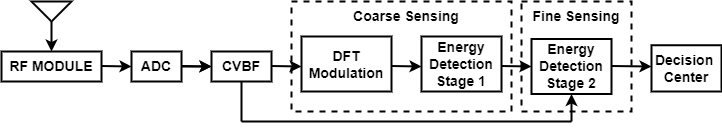}}
 \caption{Block diagram of the Proposed Method} 
 \label{m}
 \end{figure}
  \begin{figure}[h!]
 \centerline{\includegraphics[width=10 cm]{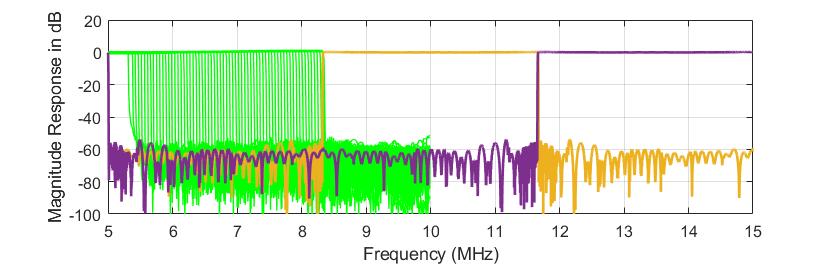}}
 \caption{64 bandwidths that can be obtained within the band.} 
 \label{m}
 \end{figure}
\begin{figure}[h!]
 \centerline{\includegraphics[width=10 cm]{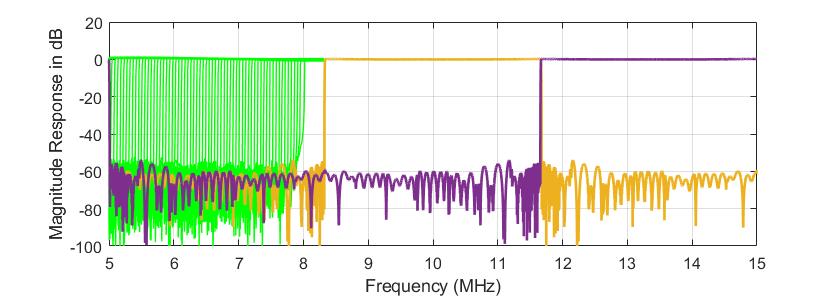}}
 \caption{64 bandwidths that can be obtained within the band from the other side of the band.} 
 \label{m}
 \end{figure}

The basic block diagram of the proposed spectrum hole detection is Figure 4. In the proposed method, there are two levels of sensing. Initial coarse sensing can be done in multiple bands in parallel. The coarse sensing is done to identify whether a particular band is occupied or not. Later fine sensing can be done in each occupied band to detect spectrum holes in the occupied bands. The bandwidth of the coarse sensing filter is determined based on two factors:\\
a) The bandwidth of the coarse sensing filter is obtained by the frequency modulation of the  continuously variable bandwidth filter design.\\
b) The modulated filters should be able to cover a range of frequency.\\
In the fine sensing stage, using the continuously variable bandwidth design, sensing is done from two sides of each band. Any arbitrary bandwidth within the band can be obtained using the continuously variable bandwidth design. Figure 5 shows 64 bandwidths that can be obtained within the band. Figure 6 shows 64 bandwidths within the band from the other side of the band. It is through low pass to high pass frequency translation, sensing is done from the other side of the band. In Figure 5 and Figure 6, a frequency range from 5 to 15 MHz is sensed. The coarse sensing filters in these Figures with bandwidth 3.33 MHz are derived from the DFT modulation of a low pass filter with bandwidth 1.665 MHz which can be readily obtained using the continuously variable bandwidth design. \\
\begin{figure}[h]
\centerline{\includegraphics[width=14 cm]{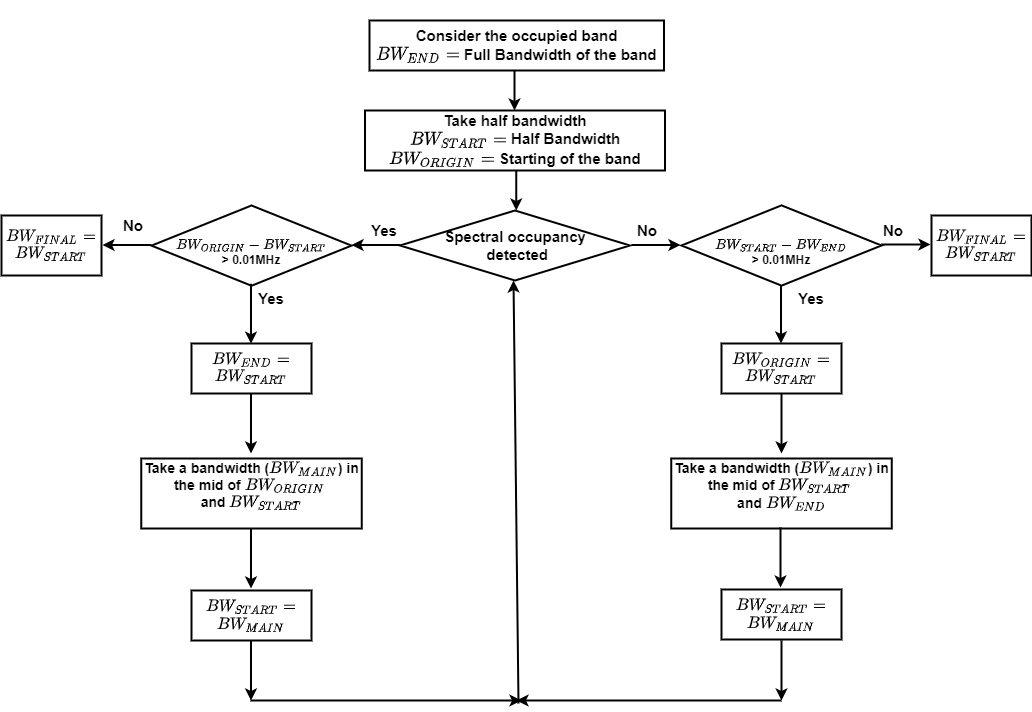}}
\caption{Flowchart of the fine sensing stage}
 \label{Fig. 9}
  \end{figure}
The user occupancy within a bandwidth can be checked using the energy detection method described in subsection 3.1. If the bandwidth is not occupied, it can be considered as a spectrum hole. Since any arbitrary bandwidth can be obtained using the continuously variable bandwidth design, spectrum holes can be detected with high resolution.\\
An efficient fine sensing can be done using the steps in the flow chart shown in Figure 7. Fine sensing is done such a way that initially the full bandwidth within the band is considered. If spectral occupancy is detected, take the half bandwidth. Check the spectral occupancy in  the half bandwidth. If the spectral occupancy is detected in the half bandwidth, take again the half bandwidth and the process continues. If the spectral occupancy is not detected in the half bandwidth, it is evident that the other half of the full bandwidth will be occupied. Hence choose a bandwidth between half bandwidth and full bandwidth and check the occupancy. If spectrum is detected in this bandwidth, take a bandwidth between the half bandwidth and the new bandwidth and if spectrum is not detected, take a bandwidth between new bandwidth and the full band width. This process is repeated until we get a proper resolution at which the spectrum begins within the band. The same process is repeated from the other side of the band. Hence with a proper resolution the spectrum beginning and ending portion within the band can be detected. So spectrum holes form two sides of the band can be detected with high resolution. Figure 7 shows a flowchart of this.
 
\subsection{Energy Detection}

Energy detection based spectrum sensing is the most widely used spectrum sensing method. In this method, prior knowledge of the primary user (PU) characteristics is not required. The binary hypothesis \(H_{0}\) and \(H_{1}\) can be defined as follows:

\begin{equation}
\begin{split}
H_{0}: &  y_{k}(n) = w(n) \\
H_{1}: &  y_{k}(n) = h_{k}(n).x_{k}(n)+w(n)
\end{split}
\end{equation}
where \( y_{k}(n)\) is the \(k^{th}\) signal received by the CVBW filter, \( x_{k}(n)\) is the PU signal, \(h_{k}(n)\) is the channel gain and \(w(n)\) is Additive White Gaussian Noise (AWGN) signal with
zero mean and variance \(\sigma_{\omega}^{2}\). \(H_{0}\) denotes the absence of the primary user signal and \(H_{1}\) denotes the presence of the primary user signal.\\
 Energy of the signal is given by
\begin{equation} 
   E =\frac{1}{N_{s}}\sum_{n=1}^{N_{s}}|y_{k}|^{2}
\end{equation}
where \(N_{s}\) is the number of samples of the signal in the available time. The performance of this scheme can be evaluated using probability of detection and probability of false alarm.
\begin{equation}
\begin{split}
P_{d} & =P(E>\lambda|H_{1})\\
P_{f} & =P(E>\lambda|H_{0})
\end{split}
\end{equation}

where \( \lambda\) is the threshold value. \(P_{d}\) and \(P_{f}\) can be expressed as:
\begin{equation} 
 P_{d}  = \int_{\lambda}^{\infty}p_{1}(x) dx
  = Q\big(\frac{\lambda-\mu_{1}}{\sigma_{1}}\big)
 \end{equation}
\begin{equation} 
  P_{f}  = \int_{\lambda}^{\infty}p_{0}(x) dx
  = Q\big(\frac{\lambda-\mu_{0}}{\sigma_{0}}\big)
\end{equation}
 Using the Central Limit Theorem, the probability density function under
hypothesis \(H_{1}\)  can be approximated by a normal distribution with mean
\(\mu_{1}= (\sigma_{x}^{2} + \sigma_{w}^{2})\) and variance \(\sigma_{1}^{2} = \frac{2}{N_{s}}(\sigma_{x}^{2} + \sigma_{w}^{2})^{2}\). Similarly \(H_{0}\)  can be approximated by a normal distribution with mean \(\mu_{0}= \sigma_{w}^{2}\) and \(\sigma_{0}^{2} = \frac{2}{N_{s}} \sigma_{w}^{4}\). Hence \(P_{d}\) and \(P_{f}\) can be expressed as:
\begin{equation} 
 P_{d} = Q\Big(\frac{\lambda- (\sigma_{x}^{2} + \sigma_{w}^{2})}{(\sigma_{x}^{2} + \sigma_{w}^{2})/{\sqrt{N_{s}/2}}}\Big)
 \end{equation}
\begin{equation} 
 P_{f} = Q\Big(\frac{\lambda- \sigma_{w}^{2}}{ \sigma_{w}^{2}/{\sqrt{N_{s}/2}}}\Big)
 \end{equation}

From equations 14 and 15 it can be analyzed that the decision threshold can be derived for either a target \(P_{d}\) or \(P_{f}\). The threshold for a constant detection rate (\(\lambda_{P_{d}}\)) and the threshold for a constant false alarm rate (\(\lambda_{P_{f}}\)) can be derived from equations 14 and 15 and are shown below.
 \begin{equation} 
  \lambda_{P_{d}}= (\sigma_{x}^{2} + \sigma_{w}^{2})\Big(1+\frac{Q^{-1}(P_{d})}{\sqrt{N_{s}/2}} \Big)
 \end{equation}

\begin{equation} 
 \lambda_{P_{f}} = \sigma_{w}^{2}\Big(1+\frac{Q^{-1}(P_{f})}{\sqrt{N_{s}/2}} \Big)
 \end{equation}

If \(P_{d}\)  is high, it means that primary users are given importance, since primary users are better protected then.
 If \(P_{f}\) is low, more chances a channel can be utilized by secondary users.
  In fixed threshold method, we are fixing the threshold either favour of \(P_{d}\) or \(P_{f}\). So a high \(P_{d}\) and low \(P_{f}\) cannot be achieved together. An adaptive threshold method is proposed in [22] where the threshold level in each sub-band is calculated with respect to the signal in that sub-band. This is used for spectrum hole detection in the proposed method. This results in a better spectrum efficiency for both primary user and secondary user. Hence spectrum holes can be detected more accurately.
   \begin{figure}[h]
\centerline{\includegraphics[width=7 cm]{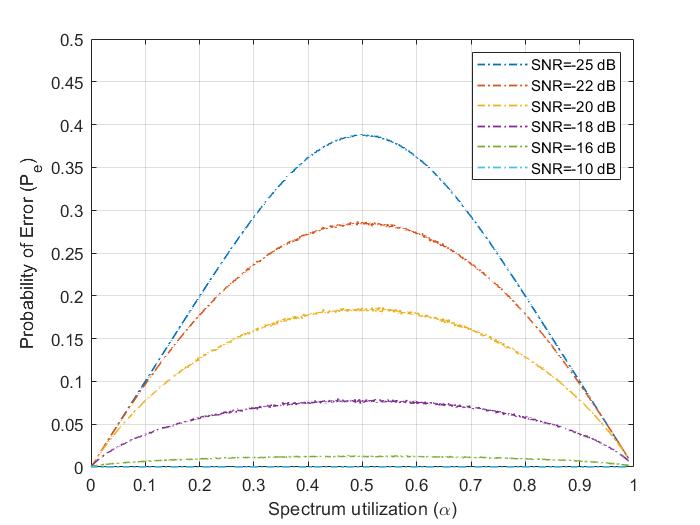}}
\caption{The probability of error versus spectrum utilization for different SNR Values}
 \label{Fig. 9}
  \end{figure}

  Adaptive threshold is found by solving [22]
  \begin{equation}
  min (P_{e}) = min( (1-\alpha) P_{f} + \alpha (1- P_{d} ) )
  \end{equation}
  where \(P_{e}\) represents the minimum probability of the error decision, \( (1- P_{d} )\)   is the probability of the missed detection, \(\alpha\) is the spectrum utilization ratio  by PU, where
\(0<\alpha<1\). A threshold is found in such a way that \(P_{f}\) is kept minimum when the spectrum is not utilized by the primary user and \((1- P_{d}) \) is kept minimum when the spectrum is utilized by the primary user. \\
 A threshold is found such a way that the probability of error is minimum which is given by

 \begin{equation}
\lambda_{adap}=\frac{1+\sqrt{1+\frac{4}{N_{s}}\big(1+\frac{2}{\gamma}\big)\ln\big(\frac{(1-\alpha)(1+\gamma)}
{\alpha}\big)}}{\sigma_{\omega}^{-2}\big(\frac{2+\gamma}{1+\gamma}\big)}
 \end{equation}
where \(\sigma_{\omega}^{2}\) is the noise variance, \(\gamma\) is the SNR and \(N_{s}\) is the sample points.
This adaptive threshold is used for spectrum hole detection in this proposed method. The probability of error versus spectrum utilization for different SNR values is shown in Figure 8.

 \section{Results and Discussion }
 In this section, the working of the proposed spectrum hole detection technique is demonstrated. To test the proposed scheme for spectrum hole detection, a continuously  variable bandwidth filter using Pascal structure and two stage FRM is first designed using PM algorithm. The following specifications are used to design the fixed filter in the continuously variable bandwidth filter design:
 \vspace{0.25 cm}
 \\
 \emph{Passband edge frequency}: \(0.14\pi\)\\
   \emph{Stopband edge frequency}: \(0.141\pi\)\\
    \emph{Maximum passband ripple}:     0.03 dB\\
   \emph{Minimum stopband attenuation}: 50 dB\\
    \vspace{0.25 cm}
 
 A Quadrature Phase Shift Keying (QPSK)
modulated signal with different center frequencies is used as the primary user signal for evaluating the spectrum sensing model. AWGN channel is considered here with a noise power known a priori. For calculating the adaptive threshold, \(\alpha\) value is chosen as 0.5. Three cases are discussed in the section. In the first case, spectrum holes in a frequency range where a single user is occupied is discussed. Spectrum hole detection while multiple users occupy the spectrum is discussed in the second case. In the third case, spectrum hole detection in a frequency range where the multiple users occupy the spectrum in close proximity is discussed. A channel of bandwidth 10 MHz is considered for evaluating the spectrum hole detection model.

\subsection{Case 1}
A Quadrature Phase Shift Keying (QPSK)
modulated signal with center frequency at 6.875 MHz and sampled at 10 MHz is used as the primary user signal in this case. Initially the coarse sensing is done by modulating the continuously variable bandwidth filter into 3 bands. The bandwidth of the coarse sensing filter is derived using the continuously variable bandwidth design through which a low pass filter is obtained with 1.665 MHz. This is modulated to obtain 3.33 MHz. Three such band pass filters will cover the entire 10 MHz.

\begin{figure}[h!]
 \centering
 \includegraphics[width= 12 cm]{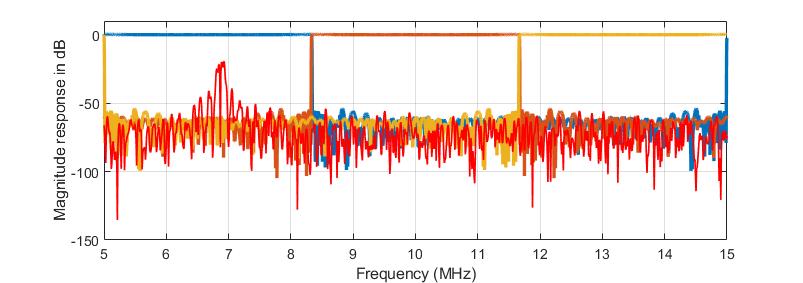}
 
  \caption{Input Signal and the coarse sensing stage}
 \label{Fig. 9}
  \end{figure}

  \begin{figure}[h!]
 \centering
 
  \includegraphics[width=12 cm]{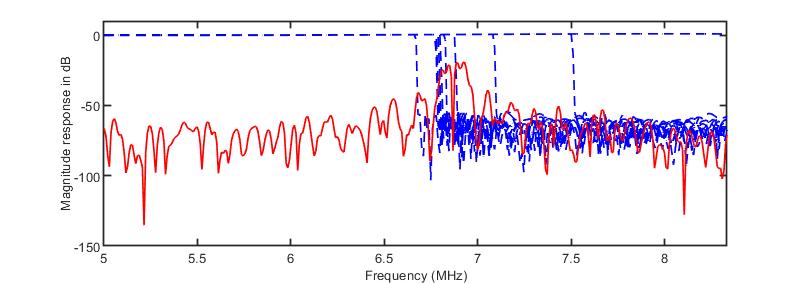}
  \includegraphics[width=12 cm]{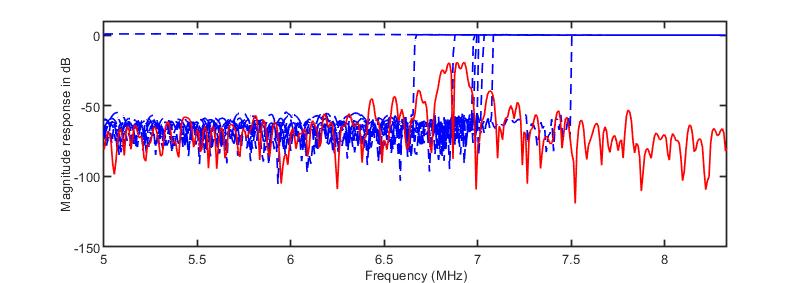}
  \caption{ Fine sensing of the occupied band}
 \label{Fig. 9}
  \end{figure}

For fine sensing the algorithm described in section 3 is used. Spectrum is detected in the frequency range  6.77 MHz to 6.9789 MHz in the first band. The unoccupied portions in the occupied band can be considered as a spectrum hole. Figure 9 shows the input signal and the coarse sensing stage. Figure 10 shows the corresponding fine sensing stage of the occupied band. Table 1 shows the energy detection associated with the coarse sensing of the spectrum holes. Table 2 shows the energy detection associated with the fine sensing of the spectrum holes. In Table 1, \(B_{1}\) is the low frequency band and \(B_{3}\) is the high frequency band. In Table 2,  \(B_{L1}\), \(B_{L2}\),  \(B_{L3}\) etc are the bands in which the energy is detected so as to find the starting of the spectrum. Here \(B_{L1}\) is the lowest frequency band and \(B_{L10}\) is the full bandwidth. The threshold, energy and occupancy corresponding to each band is shown in Table 2. \(B_{R1}\), \(B_{R2}\), \(B_{R3}\) etc are the bands in which the energy is detected so as to find the ending of the spectrum.  Here \(B_{R1}\) is the highest frequency band and \(B_{R9}\) is the full bandwidth. Corresponding threshold, energy and occupancy is shown in Table 2.

\begin{table}[h!]
\centering
\captionsetup{justification=centering}
\caption{Energy detection associated with coarse sensing }
\begin{tabular}{| l | l | l | l |}
 \hline
 Band & Threshold & Energy & occupancy\\ [1ex] 
 \hline
 \(B_{1}\) & 0.2008 & 31.6343  & Yes \\[1ex]  
\(B_{2}\) & 0.1098 & 0.0212 & No  \\[1ex] 
\(B_{3}\)  & 0.1073 & 0.0152 &  No \\[1ex]

 \hline
\end{tabular}
\label{table:1}
\end{table}

\begin{table}[h!]
\caption{Energy detection associated with fine sensing}
\begin{tabular}{| l | l | l | l | l | l | l | l |}
 \hline
 Band & Threshold & Energy & occupancy & Band & Threshold & Energy & occupancy\\ [1ex] 
 \hline
 \(B_{L1}\) & 0.1089 & 0.0189  & No  & \(B_{R1}\) & 0.1047 & 0.0093  & No \\[1ex]  
\(B_{L2}\) & 0.1318 & 0.0919 & No  & \(B_{R2}\) & 0.1169 & 0.0397 & No \\[1ex] 
\(B_{L3}\)  & 0.1424 & 0.1447 &  Yes  & \(B_{R3}\) & 0.1493 & 0.1362 & No\\[1ex] 
 \(B_{L4}\)  & 0.1566 & 0.2560 & Yes  & \(B_{R4}\) & 0.1527 & 0.1492 &  No\\[1ex] 
 \(B_{L5}\)  & 0.1742 & 0.5582 & Yes   & \(B_{R5}\) & 0.1527  & 0.2192 & Yes\\[1ex] 
 \(B_{L6}\)  & 0.1976 & 5.6810 & Yes  & \(B_{R6}\) & 0.1994 & 0.2189 &  Yes\\[1ex] 
 \(B_{L7}\)  & 0.2004 &  20.4982 & Yes & \(B_{R7}\) & 0.2007 & 10.9126  & Yes\\[1ex] 
 \(B_{L8}\)  & 0.2008 & 31.4512  & Yes  & \(B_{R8}\) & 0.2007 & 28.3152  & Yes\\[1ex] 
 \(B_{L9}\)  & 0.2008  & 31.2684 & Yes  & \(B_{R9}\) & 0.2007 & 28.9599 & Yes\\[1ex] 
  \(B_{L10}\) & 0.2008  & 31.6343   & Yes  & - & - & -  & -\\[1ex] 
 
 \hline
\end{tabular}
\label{table:1}
\end{table}
 \subsection{Case 2}
 
A Quadrature Phase Shift Keying (QPSK)
modulated signal with center frequencies at 6.875 MHz and  12.75 MHz and sampled at 10 MHz is used as the primary user signal in this case. By coarse sensing, spectrum is detected in the first band and third band. Then fine sensing is done in each of these bands separately. Fine sensing in first band is the same as in case 1. The coarse sensing stage and the fine sensing in the third band is shown in Figure 11.

\begin{figure}[h!]
 \centering
 
 \includegraphics[width=12 cm]{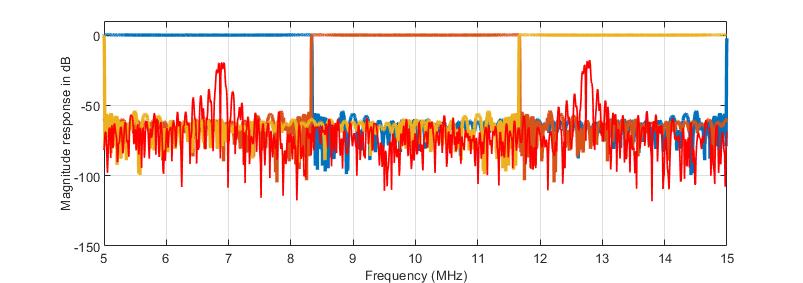}
 \includegraphics[width=12 cm]{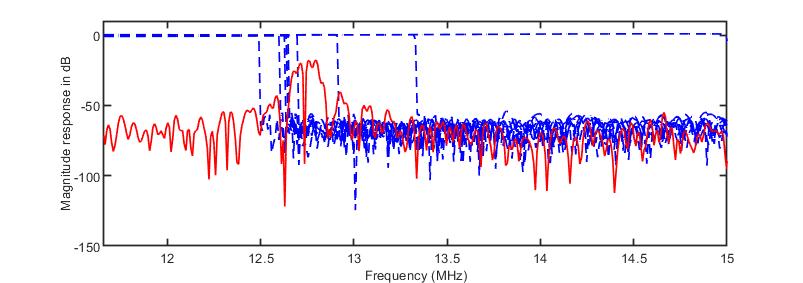}
 \includegraphics[width=12 cm]{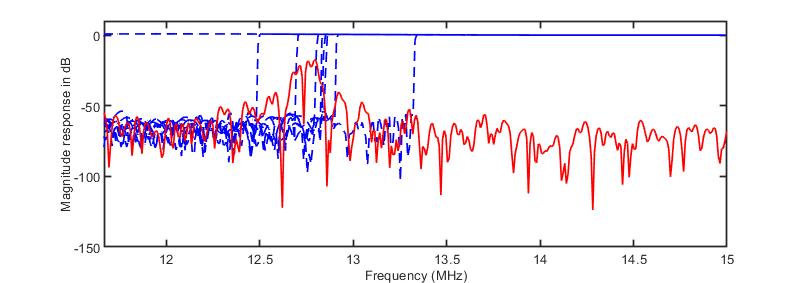}
 \caption{Spectrum hole detection with multiple primary users}
 \label{Fig. 9}
  \end{figure}
 
\subsection{Case 3}
When multiple primary users are there in close proximity, chances are there that the spectrum hole in between two primary users will be missed. In order to avoid such situation, when a band in coarse sensing stage is having high energy, chances are there that multiple bands are there within a band. In such cases, number of bands of the coarse sensing stage can be increased. Here in this case 6 bands are taken instead of 3 bands. This is shown in Figure 12. Since using continuously variable bandwidth filter can give multiple bandwidth filter as the prototype filter, increasing the number of bands in a coarse sensing stage does not demand the design of a new prototype filter. Only the complexity for the modulation of the filter will be increased. In this case a Quadrature Phase Shift Keying (QPSK)
modulated signal with center frequencies at 9 MHz and  11 MHz and sampled at 10 MHz is used as the primary user signal. Fine sensing is done similar to the cases 1 and 2.
 
  \begin{figure}[h!]
 \centering
 \includegraphics[width=12 cm]{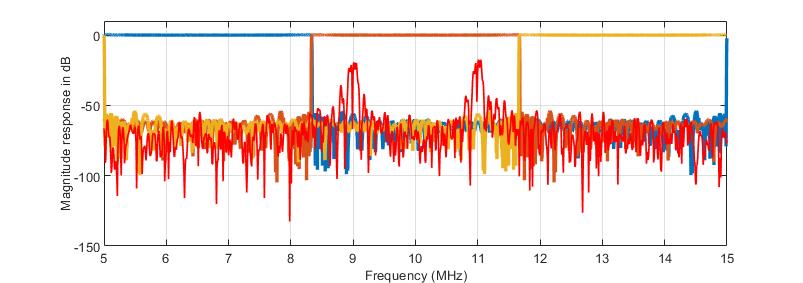}
 \includegraphics[width=12 cm]{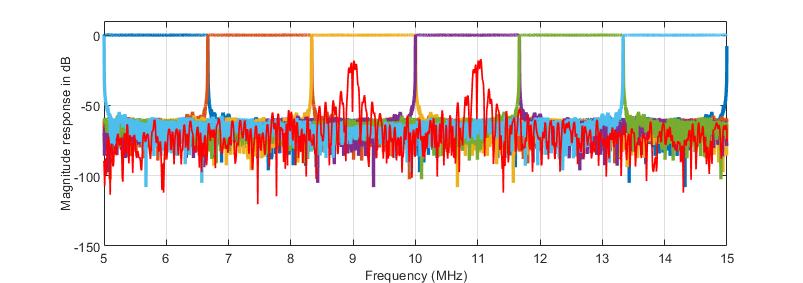}
 \caption{Spectrum hole detection with multiple primary users in close proximity}
 \label{Fig. 9}
 \end{figure}
 \begin{table}[h]
\centering
\caption{Hardware Complexity Comparison}
\begin{tabular}{| l | l | l | l |}
 \hline
 Design Method & Transition & Multipliers in  & Total \\ 
 &Width& Prototype filter& Multipliers\\
 \hline
 CMFB based [14] &\( 0.0025\pi\) & 1396  & 2226 \\[1ex]  
Farrow Structure [16] & \( 0.0025\pi\) & 700& 888
 \\[1ex] 
Modified FRM based [15] & \( 0.0025\pi\) &  145 & 465 \\[1ex] 
Proposed Method & \( 0.001\pi\) & 228 & 266\\[1ex] 
 \hline
\end{tabular}
\end{table}

 \section{Hardware Complexity Calculation}
Table 3 shows the hardware complexity comparison of the proposed scheme with the state of the art. Hardware complexity is calculated by considering both the prototype filter and multipliers for modulation. The complexity of determining the threshold level to perform test statistics for the spectrum hole detection is excluded in the computation as it's not much significant when compared to the complexity of the prototype  filter and the modulation. The hardware complexity is given by the equation:
\begin{equation}
\mu_{Total}=\mu_{Prototype}+\mu_{modulation}+\mu_{HPF}
\end{equation}
\(\mu_{Total}\) is the total number of multipliers, \(\mu_{Prototype}\) is the multipliers of the prototype filter, \(\mu_{modulation}\) is the multipliers for modulation and \(\mu_{HPF}\) is the number of multipliers required for generating the high pass filter from low pass filter. Since DFT modulation is used, number of multipliers for modulation is
\begin{equation}
\mu_{modulation}=N_{c}^{2}
\end{equation}
\(N_{c}\) is the number of coarse sensing bands. Since in the proposed design's case, we are not restricting \(N_{c}\) to be a power of two, FFT (Fast Fourier Transform) is not guaranteed for DFT computation. Hence equation 21 is used for calculating number of multipliers for modulation.
Since the fixed filter is a low pass filter, to sense from either side, high pass filter is used. Low pass filter to high pass filter conversion can be considered as a case with number of channels, M=2. Hence the number of multipliers for this conversion is 
\begin{equation}
\mu_{HPF}=Mlog_{2}M
\end{equation}
This gives \(\mu_{HPF}\) as 2.\\
The case where \(N_{c}\) is taken as 6 and the prototype filter is designed as per section 4 is taken for finding the number of multipliers in the proposed scheme. This is shown in Table 3.
\begin{equation}
228+36+2=266
\end{equation}

 \section{Hardware Implementation and Complexity Evaluation}
 The complexity of implementation, power consumption and area of the proposed spectral hole detection scheme is evaluated by synthesizing and implementing the prototype continuously variable bandwidth filter in both ASIC (Application Specific Integrated Circuit) and FPGA  (Field Programmable Gate Array) platforms. 14-bit quantized prototype filter coefficients synthesized and implemented in Xilinx Artix 7 AC 701 (XC7A200T-2FBG676C) FPGA using Vivado design suit and the resource consumptions are evaluated at 20 MHz frequency.  The results are shown in Table 4.

 \begin{table}[h!]
\centering
\caption{FPGA Synthesis results}
\begin{tabular}{| l | l | l |}
 \hline
  & Slice LUTS & FFs   \\ 

 \hline
 Available resources & 1,34,600 &  2,69,200  \\
 & &  \\
 \hline
 Prototype filter in [23] & 50078 & 56636 \\
  & &  \\
  \hline
Prototype filter in the proposed method & 29621 & 33254\\
\hline
\end{tabular}
\label{table:1}
\end{table}

\begin{table}[h!]
\centering
\caption{ASIC Synthesis results}
\begin{tabular}{| l | l | l | l | }
 \hline
  & No. of cells & Area & Power  \\ 
  & & (\(\mu m^{2}\)) & (mW) \\
\hline
  Prototype filter in [23]  & 1670693  & 14681196 &  175.186  \\ 
 & & & \\
  
 \hline
  Prototype filter & 1214170  & 10554162 & 123.447  \\ 
in the proposed method& & &   \\
\hline
\end{tabular}
\label{table:1}
\end{table}
To evaluate the performance of the proposed spectrum hole detection scheme in the ASIC platform, 14-bit quantized prototype filter coefficients  synthesized in 90 nm technology ‘fast’
library using Cadence Genus synthesis tool. The results are shown in Table 5.
 \section{Conclusion}
A high resolution spectrum hole detection method is proposed in this paper. A continuously  varying bandwidth filter designed using Pascal structure and two stage FRM is deployed for this purpose. In the proposed method, the coarse sensing is done using frequency modulation of the continuously variable bandwidth filter for detecting the occupied bands. Fine sensing is done by varying the bandwidth within an occupied band for detecting the spectrum holes. Adaptive threshold based energy detection is used in the proposed method for spectrum sensing. This results in a better spectrum efficiency for both primary user and secondary user. The hardware complexity of the proposed spectrum hole detection method outperforms the prior techniques in the literature.

\end{document}